\documentclass[12pt]{iopart}
\pdfoutput=1 
\usepackage{cite}
\usepackage{xypic}
\usepackage{graphicx}
\usepackage{amsfonts}
\usepackage{mathrsfs}
\usepackage{amssymb}
\input{epsf.sty}
\catcode `\@=11
\def\numberbysection{\@addtoreset{equation}{section}
         \renewcommand{\theequation}{\thesection.\arabic{equation}}}
\numberbysection
\def\subsubsection{\@startsection{subsubsection}{3}%
  \normalparindent{.5\linespacing\@plus.7\linespacing}{-.5em}%
  {\normalfont\bfseries}}
\def\R{{\mathbb R}}

\def\Z{{\mathbb Z}}
\def\T{{\mathbb T}}

\def\H{{\mathscr H}}

\def\B{{\mathscr B}}

\def\BTheta{\B_{\Theta}}

\def\bgp{\bar\Gamma_+}

\def\Ttheta{\T^3_{\Theta}}

\begin{document}

\title{The noncommutative geometry of wire networks from triply periodic surfaces
}

\author
[Ralph M.\ Kaufmann]{Ralph M.\ Kaufmann}

\address{Department of Mathematics, Purdue University, 
 West Lafayette, IN 47907}
\ead{rkaufman@math.purdue.edu}
\author
[Sergei Khlebnikov]{Sergei Khlebnikov}

\address{Department of  Physics, Purdue University,
 West Lafayette, IN 47907}
\ead{skhleb@physics.purdue.edu}
\author
[Birgit Kaufmann]{Birgit Wehefritz--Kaufmann}

\address{Department of Mathematics and Department of Physics, Purdue University,
 West Lafayette, IN 47907}
\ead{ebkaufma@math.purdue.edu}
\submitto{\JPA}
\maketitle

\begin{abstract}
We study wire networks that are the complements of triply periodic minimal surfaces.
Here we consider the P, D, G  surfaces which are exactly the cases in which the corresponding graphs are symmetric and
self-dual. Our approach is using the Harper Hamiltonian in a constant magnetic field as set forth in \cite{B,BE,KKWK}. We treat this
system with the methods of noncommutative geometry and obtain a classification for all the $C^*$ geometries
that appear.
\end{abstract}

\section*{Introduction} It is well known that the only triply periodic minimal surfaces whose complements are given by symmetric and self-dual graphs  are the P, D and G surfaces , see e.g. \cite{Anderson}. While the P and D surfaces were already discovered by Schwarz in 1830 \cite{Schwarz}, it took until 1970 for the G surface to be discovered by Alan Schoen \cite{Schoen}. In real situations these surfaces appear as the boundary between phases. 
We will concentrate on the complement of these surfaces which consists of two components or channels. For the P, D and G surfaces
these two channels have the same underlying skeletal graph onto which they retract. This graph carries all the
homotopical information, such as the K-theory. Figures \ref{Pgraph}, \ref{Dgraph} and \ref{Ggraph} show one channel and its skeletal graph for the respective cases.

The guiding physical motivation 
for this study is that, when the boundary has a finite thickness
(as it always does in real materials), the complement still forms two channels of 
a nanoporous structure. These channels can be filled with a (semi)conductor,
forming a nanowire network of potential interest in applications.  
Indeed, for
the G surface, or rather the double Gyroid, this has been achieved \cite{Hillhouse}. Each channel is composed of
approximately cylindrical segments joined together at triple junctions. Numerical simulations of a simple wave equation \cite{Khlebnikov&Hillhouse}
have shown that the lowest-energy wavefunctions are supported primarily on the junctions. Thus, one may expect to reproduce the low-energy
end of the spectrum by using the tight-binding approximation, in which
the junctions are replaced by the vertices, and the segments connecting them
by the edges, of a graph. Mathematically speaking, this means  that each component of the  complement of
a G surface is indeed contracted onto its graph. 

Of particular interest is the behavior of periodic nanoporous materials in an external
magnetic field, specifically, the questions of existence and number of any additional
gaps in the spectrum the field may produce. Such gaps would be a 3-dimensional analog of Hofstadter's butterfly \cite{Hofstadter}. Note that the materials in question are ``supercrystals,'' whose lattice constants far exceed atomic
dimensions. (For instance, for the double gyroid of \cite{Hillhouse}, the lattice constant is of order 20 nm.) As a result, the magnetic flux through the unit cell may be a sizable
fraction of the flux quantum for realistic magnetic fields, opening the possibility
of an experimental study of the additional gap structure. We would like to understand this structure from the point of view of non-commutative geometry, in parallel with the earlier studies of the quantum Hall effect \cite{BE}.

In our previous article \cite{KKWK}, we gave a general approach for such wire systems treated as graphs with a given translational symmetry group. The relevant result of this analysis was that  for a constant magnetic field the relevant $C^*$ algebra $\B$ generated by magnetic translation operators and the Harper Hamiltonian has a faithful matrix representation as a subalgebra of a matrix algebra of a noncommuative torus. More precisely, let $n$ be the dimension of the ambient space,  $k$ be the number of sites in a primitive cell and $B=2\pi \Theta$ be the magnetic field expressed as a 2--form. Then $\B$ embeds into $ M_k(\T^n_{\Theta})$,
where $\T^n_{\Theta}$ is the noncommutative n--torus with parameter $\Theta$ and $M_k(\T^n_{\Theta})$ is the $C^{\star}$ algebra of $k \times k$ matrices with entries in $\T^n_{\Theta}$. One expects
that generically, that is if all entries of $\Theta$ are irrational, the algebra $\B$ is the full matrix algebra and thus is Morita equivalent to $\T^n_{\Theta}$ itself. At rational points there is no such expectation. An interesting question is to classify
the points at which the algebra is a proper subalgebra, as they should have special physical properties.

Applying this general theory, we will focus on the case $n=3$ and the graphs arising from the P, D and G surfaces.  
The classification of special points and their $C^*$ algebras for the $G$ surface was one of the main aims of \cite{KKWK}.
We will review those results here giving a concise statement of the main results. The $P$ surface is much simpler since in this case $k=1$. The D case has not yet been considered before, and we give the complete entirely new calculation here.

\section{General background}

The general setup is following the noncommutative approach we call Connes-Bellissard-Harper approach \cite{B, MM, Connes, Harper}. We start by considering a $C^*$ algebra $\B$ which is the smallest algebra containing the Hamiltonian and the symmetries.

The standard choice of the Hamiltonian is the Harper Hamiltonian \cite{Harper}. This acts on the
 Hilbert space $\H=\ell^2(\Lambda)$ where $\Lambda$ are
the vertices of the graph. Physically, this corresponds to using the tight-binding approximation and Peierls
substitution \cite{PST}. If
we turn on a magnetic field this procedure expresses the Hamiltonian in terms of a sum of
magnetic translation or Wannier \cite{Harper} operators. 
In the general setting the magnetic field will be given by a two form on $\R^n$ which in $\R^3$ restricts
to the familiar vector field $B$. We will concentrate on the case of a constant magnetic field 
$B=2\pi \theta_{ij}dx^idx^j$ where $\Theta=(\theta)_{ij}$ is the skew--symmetric  matrix of a skew--symmetric 2-tensor.

We will now  describe our setup in more detail. 
Fix $\Gamma\subset \R^n$ to be a connected embedded graph whose edges are line segments. We denote by 
$L$ a (maximal) translational symmetry group of $\Gamma$, s.t.
$\bar \Gamma=\Gamma/L$ is finite. Here a translational symmetry group is a group isomorphic
to a free Abelian group of rank $n$ which acts by translations on $\R^n$ leaving $\Gamma$ invariant.
 Let $\pi:\Gamma\to \bar \Gamma$  be the projection. The vertices of $\bar\Gamma$ are the vertices
in a primitive cell, but the graph $\bar\Gamma$ is just an abstract graph\footnote{The graph  $\bar\Gamma$ is naturally
embedded in the torus $\R^n/\Gamma$, but not in $\R^n$ itself.}.
Let $\Lambda$ be the set of vertices of $\Gamma$, $\bar\Lambda$ the set of vertices of $\bar\Gamma$, and denote by 
$T$ the (free Abelian) subgroup of $\R^n$ generated by the {\em edge vectors}. 

Notice that $L\subset T$, but in general this inclusion is strict. 
On $\H$ a magnetic translation by a vector ${\bf e}$ of $L$  is represented by a unitary operator $U_{\bf e}$, while a translation by a vector
in $T$ only gives rise to a partial isometry.
To see this, we decompose the Hilbert space 
$\H=\bigoplus_{v\in \bar\Lambda} \H_{v}$ where $\H_v=l^2(\pi^{-1}(v))$. Then a translation by $e\in T$ which
goes from $w$ to $v$ will act as  $U_{e}:\H_v\to \H_w$.
\footnote{This is assuming
the standard action for magnetic translation operators.}
In this formalism, the Hamiltonian is represented by a sum of partial isometries. 

As it is defined $\B$ is a $C^*$ sub-algebra of the operators on $\H$. 
In order to calculate the algebra $\B$ more explicitly, we wish to define a matrix representation of it. 
For this one fixes a rooted spanning tree.
A  spanning tree is a subtree of the graph, which contains all vertices. Being rooted means that
one vertex is distinguished.

Our main theorem which allows us to do explicit computations is then:

{\bf Theorem.} \cite{KKWK}
For $\Gamma$, $L$ as above and a fixed $B$ given by $2\pi \Theta$,  
fixing a choice of rooted spanning tree  for $\bar \Gamma$, an order of the vertices of $\bar \Gamma$
and a basis for $L$ defines a  faithful
matrix representation of $\B$ which is a sub--$C^*$--algebra $\BTheta$ of the $C^*$
algebra $M_{|V(\bar\Gamma)|}(\T^{n}_{\Theta})$, where $\T^{n}_{\Theta}$ is the noncommutative torus.

{\bf Consequence.} From this it follows, that if $\Theta$ is rational then the spectrum of $H$ has finitely many gaps.
Moreover the maximal number is determined by the entries of $\Theta$. 

In the case of the square lattice, this gives rise to the Hofstadter butterfly \cite{Hofstadter}. Hence our result can be viewed as 
a generalization to the lattices of our setup.

In the above  theorem, the translations of $L$ are what gives rise to the noncommutative torus. In particular
each fixed basis element $e_i$ of $L$ gives rise to a unitary diagonal operator valued matrix $\rho(U_i)$. These
matrices satisfy the commutation relations  $\rho(U_i)\rho(U_j)=e^{2\pi\theta_{ij}} \rho(U_j)\rho(U_i)$
and hence give a representation of $\T^{n}_{\Theta}$ which is the $C^*$ algebra spanned by $n$ independent
unitary operators $U_i$ satisfying the commutation relations $U_iU_j=e^{2\pi\theta_{ij}} U_jU_i$.
In  the matrix representation of the
 Hamiltonian, each partial isometry which corresponds to the summand of the Hamiltonian describing the translation 
  along the edge joining  the vertex
$k$ to the vertex $l$ gives rise to a $\T^{n}_{\Theta}$ valued 
matrix entry in the $(l,k)$-th position.
 
Notice that there are two incarnations of the Harper Hamiltonian, the first, which we will simply call 
the Harper Hamiltonian is the operator acting on $l^2(\Lambda)$. The second one is its representation in the matrix ring $M_k(T^n_{\Theta})$ which 
we call the matrix Harper Hamiltonian.

{\bf Associated (non)-commutative geometries. }
On general grounds we expect three types of different possible phenomenologies according to whether (a) $\Theta=0$ and
there is no magnetic field, (b) $\Theta$
is generic (i.e.\ all entries are irrational), (c) $\Theta$ contains rational entries. 

If $\Theta=0$ then the $C^*$ algebra is a unital commutative and by the Gelfand-Naimark Theorem
it is isomorphic to the $C^*$ algebra $C(X)$ of continuous $\mathbb C$ valued functions on
a compact Hausdorff space $X$. Thus starting with $\Gamma$ and $T$ in $\R^3$ we get a new
geometry $X$. Here the base $T^n$ is given by the possible exponential values of momenta in the basis directions
of $L$. \footnote{Notice these are {\em not} the momenta along the x,y,z axis.} The cover can then be interpreted as the different Eigenvalues of $H$.  Let us call a point
non--degenerate if $H$ at theses fixed momenta has $n$ distinct Eigenvalues. Since this is an
open condition, we get that  if there is one point which is
non--degenerate then that this is generically the case. In general, we showed \cite{KKWK}

{\bf Theorem.} The space  $X$ is a branched cover of the torus $T^n=S^1\times \dots \times S^1$
 ramified over the locus where $H$ has  degenerate Eigenvalues .

When $\Theta$ is {\em generi}c, it is known $\T_{\Theta}^n$ is simple, 
which means that it has no two sided proper ideals. So, we {\bf expect}
$\BTheta=M_{|V(\bar\Gamma)|}(\T^{n}_{\Theta})$ which is Morita equivalent to $\T^{n}_{\Theta}$.
That is the noncommutative geometry of $\Gamma$ in the magnetic field $B$ is given by the noncommutative
torus.  This is not a proof, however, and it has to be checked in each case.

When $\Theta$ {\em contains rational entries}, there is {\bf no expectation} and in a sense
this is the most interesting case. It can happen that the resulting 
algebra $\BTheta$ is (i) commutative, this corresponds to special commensurabilities, (ii) that it is again 
the full matrix algebra or (iii) that it is a proper subalgebra of the matrix algebra.

In the next section, we will analyze the three cases of the P, D and G  wire networks explicitly.
In the case of $\R^3$ the skew--symmetric bilinear form $\Theta$ given by $B$ takes on the familiar form 
$$\Theta(v,w)=
\frac{1}{2\pi}B \cdot (v\times w)$$.

\section{Specific results for the cubic (P) case}
\begin{figure}
\begin{center}
\includegraphics[height=5cm]{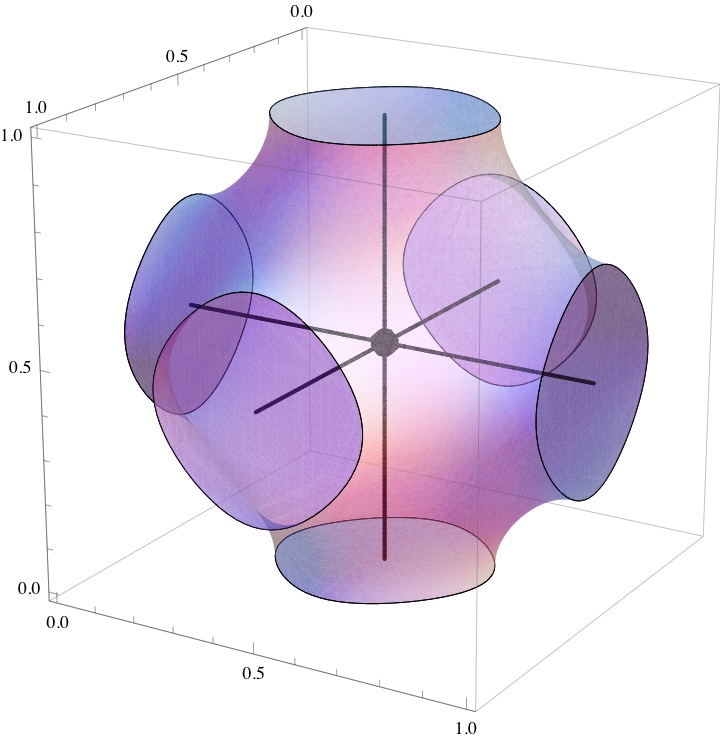}
\caption{One channel of the P surface and its skeletal graph. This and Figures 2 and 5 were obtained using the level surface approximation for the corresponding minimal surfaces \cite{levelsurface}.}
\label{Pgraph}
\end{center}
\end{figure}

The P surface has a complement which has two connected components each of which can be retracted to the simple cubical graph whose vertices are the
integer lattice $\Z^3\subset \R^3$. The translational group is again $\Z^3$ in this embedding as shown in Figure \ref{Pgraph}, so it reduces to the case of a Bravais lattice which 
we treated  already in \cite{KKWK}. Let us review some of the details.
The graph $\bar \Gamma$ is the graph with one point and three loops, so $n=1$. 
Fixing the standard basis $e_1,e_2,e_3$ of $\Z^3$,
we get the operators $U_1,U_2,U_3$, which generate $T^3_{\Theta}$ and the Hamiltonian is simply $H=\sum_i (U_i+U_i^*)$.
If $\Theta\neq 0$ then $\BTheta$ is simply the noncommutative torus and if $\Theta=0$ then 
this is the $C^*$ algebra of $T^3$.

\section{The diamond lattice (D) case}
\begin{figure}
\begin{center}
\includegraphics[height=5cm]{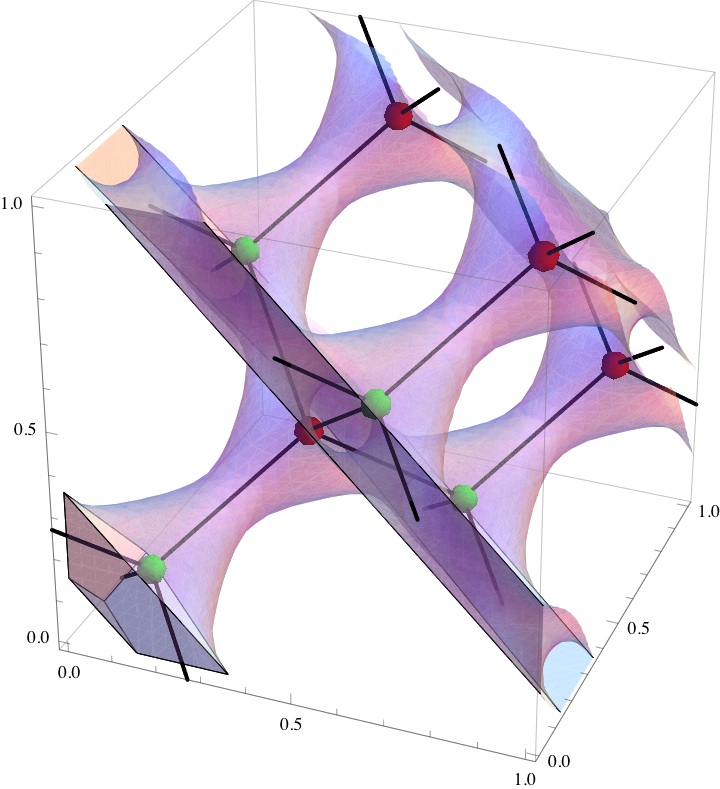}
\caption{One channel of the diamond surface and its skeletal graph. The red and green dots refer to the vertices of the two interlaced fcc lattices}
\label{Dgraph}
\end{center}
\end{figure}

\begin{figure}
\begin{center}
\includegraphics[height=3cm]{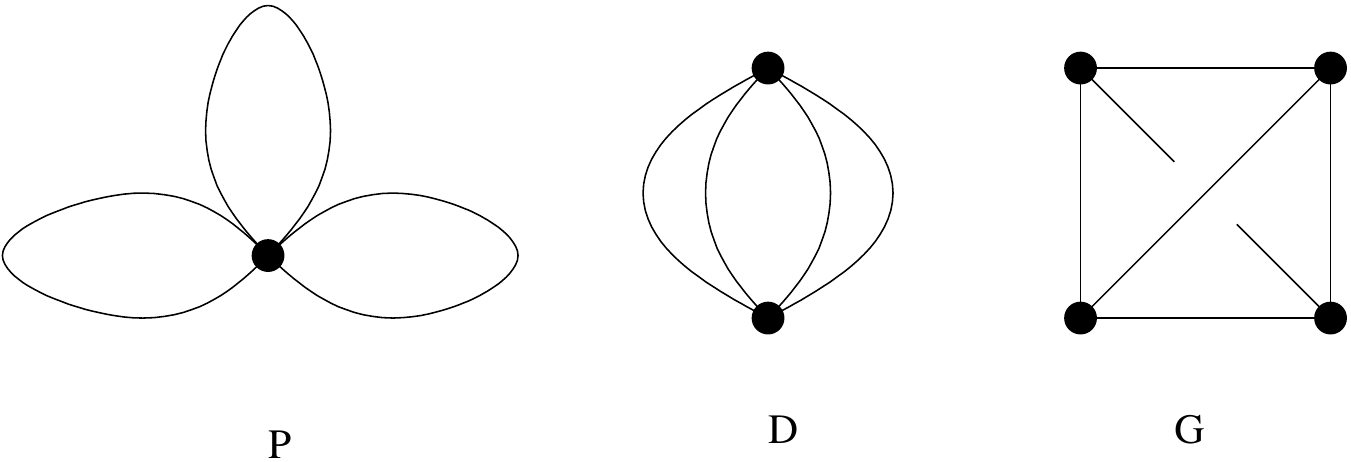}
\caption{The quotient graphs for the cubic, diamond and gyroid lattices}
\label{graphfig}
\end{center}
\end{figure}

The D surface has a complement consisting of two channels each of which can be retracted to the diamond lattice $\Gamma_{\diamond}$.  The diamond lattice
is given by two copies of the fcc lattice, 
where the second fcc is  the shift by $\frac{1}{4}(1,1,1)$ of the standard fcc lattice, see Figure \ref{Dgraph}. The edges
are nearest neighbor edges. 
 The symmetry group is $Fd\bar3m$. 
In the diamond lattice case, we have 2 vertices in the primitive cell. The quotient graph $\Gamma_{\diamond}/fcc$ is the graph with 2 vertices and 4 edges connecting them, see Figure \ref{graphfig}. The edges
correspond to the 4 vectors to the center of a tetrahedron centered at $(0,0,0)$.

$$e_1=\frac{1}{4}(1,1,1), e_2=\frac{1}{4}(-1,-1,1), e_3=\frac{1}{4}(-1,1,-1), e_4=\frac{1}{4}(1,-1,-1)$$

These vectors satisfy $\sum_i e_i=0$. We parameterize the $B$ field by fixing  the values of the skew--symmetric bilinear form  $\Theta$ on the basis elements $(-e_1,e_2,e_3)$ as follows:

\begin{equation*}
{\Theta} (-e_1,e_2)=\varphi_1\quad
{\Theta} (-e_1,e_3)=\varphi_2\quad
{\Theta} (e_2,e_3)=\varphi_3
\end{equation*}
 Our results will depend on the phases:
\begin{equation}
\chi_i=e^{i \varphi_i}\;\mbox{for}\; i=1,2,3
\end{equation}

The Harper Hamiltonian according to the construction of \cite{KKWK}  in terms of the partial isometries reads

$$
\left(
\begin{array}{cc}
0 & U^*_{e_1}+U^*_{e_2}+U^*_{e_3}+U^*_{e_4}\\
U_{e_1}+U_{e_2}+U_{e_3}+U_{e_4} & 0
\end{array}
\right)
$$

Before we can write down the matrix Harper Hamiltonian, we have to fix some data and notations.
The three edges of the tetrahedron incident to one point are $f_2=\frac{1}{2}(-1,-1,0),f_3=\frac{1}{2}(-1,0,-1),f_4=\frac{1}{2}(0,-1,-1)$. 
The translation operators along those edges fulfill the following commutation relations:
\begin{equation}
U_{f_i} U_{f_j} = e^{ 2 \pi i {\Theta}(f_i,f_j)}U_{f_j} U_{f_i}
\end{equation}

We set $U=\chi_1U_{f_2}$, $V=\chi_2U_{f_3}$ and $W=\bar \chi_1\bar \chi_2U_{f_4}$.

These operators span a $\T^3_{\Theta}$:
\begin{equation}
U V = q_1 V U \quad
UW  = q_2 WU\quad
VW = q_3 WV
\end{equation}
where the $q_i$ expressed  in terms of the $\chi_i$ are:
\begin{equation}
q_1=\bar{\chi_1}^2 \chi_2^2 \chi_3^2 \quad
q_2=\bar{\chi_1}^6 \bar{\chi_2}^2 \bar{\chi_3}^2 \quad
q_3=\bar{\chi_1}^2\bar{ \chi_2}^6 \chi_3^2 
\end{equation}
Vice versa, fixing the values of the $q_i$ fixes the $\chi_i$ up to eighth roots of unity:
\begin{equation}
\chi_1^8=\bar{q}_1 \bar{q}_2 \quad
\chi_2^8=q_1 \bar{q}_3  \quad
\chi_3^8=q_1^2\bar{q}_2 {q}_3  
\end{equation}
Other useful relations are $ q_2 \bar{q}_3= \bar{\chi}_1^4 \chi_2^4 \bar{\chi}_3^4$ and $q_2 q_3 =\bar{\chi}_1^8 \bar{\chi}_2^8$.

Using the $e_1$ edge as the spanning tree with the root being the vertex that corresponds to $\pi(0,0,0)$, we get that
the embedding representation $\rho$ of $\T^3_{\Theta}$ into $M_2(\T^3_{\Theta})$ defined by the action of $L$ is
given by
\begin{eqnarray}
\rho(U)&=&diag(U, \chi^2_1U),\;\; \rho(V)=diag(V,\chi^2_2 V)\nonumber\\
\rho(W)&=&diag(W,\bar\chi^2_1\bar\chi^2_2 W).
\end{eqnarray}
And  the matrix Harper Hamiltonian is
$$
H=
\left(
\begin{array}{cc}
0&1+U^*+V^*+W^*\\
1+U+V+W&0
\end{array}
\right)
$$

\subsection{The commutative case}
In this case, we see that the algebra $\BTheta$ is a subalgebra of $M_2(C(T^3))$, where $C(T^3)$ is the $C^*$ algebra of complex functions on the torus $T^3$.

The space $X$ corresponding to the commutative $C^*$ algebra  is a ramified cover of $T^3$ which is generically $2:1$. The branching locus is given by the degenerate points. These are computed by:

$$
det(H-\lambda id)= \lambda^2-(1+U+V+W)(1+U+V+W)^*
$$
There are degenerate Eigenvalues of $H$ on a point of $T^3$ which corresponds to the 
character $\chi: \T^3\to {\mathbb C}$, given by evaluating at that point, if the following equations are satisfied:
Set $e^{i\phi_1}=z_1=\chi(U),e^{i\phi_2}=z_2=\chi(V),e^{i\phi_3}=z_3=\chi(W), \in S^1\subset {\mathbb C}$ then the square root has only one value $0$ if
$$
1+z_1+z_2+z_3=0
$$
We calculate
$$
-z_1=1+z_2+z_3 
$$
$$
1=z_1\bar z_1=1+z_2\bar z_2+z_3\bar z_3+z_2+\bar z_2 +z_3+\bar z_3+z_2\bar z_3+\bar z_2z_3
$$
multiplying by $z_2z_3$
$$
0=2z_2z_3+z_2^2z_3+z_3+z_2z_3^2+z_2+z_2^2+z_3^2=(z_2+z_3)(z_2+z_3+1+z_2z_3)
$$
This gives the solution $z_2=-z_3, z_1=-1$ or $z_2(z_3+1)=-(z_3+1)$. The latter equation has the solutions $z_3=-1,z_1=-z_2$ and $z_2=-1, z_3=-z_1$.

\begin{figure}
\begin{center}
\includegraphics[height=5cm]{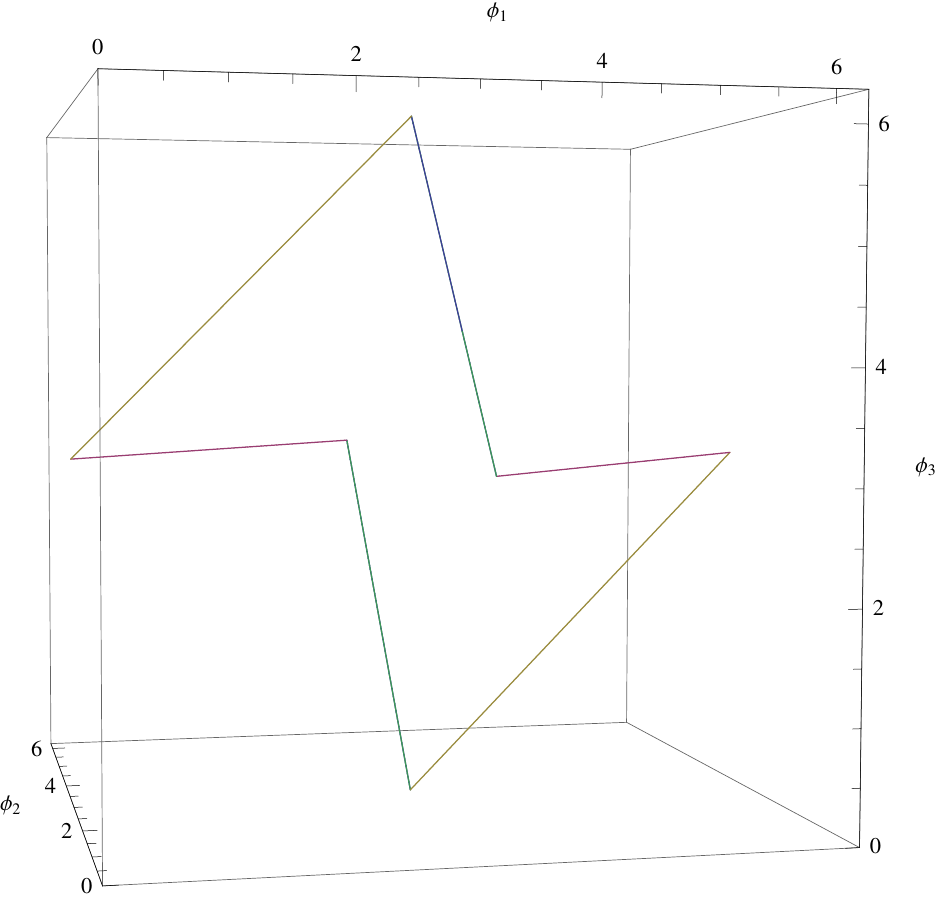}
\caption{Commutative case for the D surface- ramification locus on $T^3$ depicted as the cube with periodic boundaries}
\label{solcomm}
\end{center}
\end{figure}

{\bf Cover of $T^3$ defined by the $D$ wire network.} 
We see that the space $X$ defined by $\B$ in the commutative case is a generically 2--fold cover of the 3--torus $T^3$ where
 the ramification is along three circles on $T^3$ given by the equations $\phi_i=\pi, \phi_j\equiv \phi_k+\pi\; \mbox{mod}\; 2\pi$ with $\{i,j,k\}=\{1,2,3\}$.
 They are shown in Figure \ref{solcomm}, where the cube has to be taken with periodic boundaries. Therefore the intersection points on opposite faces of the cube  are identified 
 and the six lines form three circles which pairwise touch at a point.

\subsection{The non--commutative case}

In the following we would like to characterize the algebra $\BTheta$ for general values of the magnetic field.
The results will split into cases according to the values of the parameters $q_i$ and $\chi_i$.

In a first step, set $ X_1= H -\rho(\bar\chi^2_1 U) H \rho(U^*)$.

$$
X_1=
\left(
\begin{array}{cc}
0& (1-\bar{\chi_1}^4)(1+U^*)+(1-\bar{\chi_1}^4\bar{q_1})V^*+(1-\bar{\chi_1}^4\bar{q_2})W^*\\
(1-q_1) V +(1-q_2) W&0
\end{array}
\right)
$$
Now, set $X_2 =  X_1 -\rho(\bar\chi_2^2V) X_1 \rho(V^*)$ and $X_3= X_2 -\rho(\chi_1^2\chi_2^2) X_2 \rho(U_{f_4}^*)$.
We obtain

\begin{equation}
\label{xmatrix}
X_3=
\left(
\begin{array}{cc}
0& a 1+ b U^*+c V^*+d W^*\\
0&0
\end{array}
\right)
\end{equation}
with 

$$a=(1-\chi_1^4 \chi_2^4)(1-\bar{\chi_2}^4)(1-\bar{\chi_1}^4), \quad  b=(1-\chi_1^4 \chi_2^4 \,q_2)(1-\bar{\chi_2}^4q_1)(1-\bar{\chi_1}^4)
$$
$$
c= (1-\chi_1^4 \chi_2^4\, q_3 )(1-\bar{\chi_2}^4)(1-\bar{\chi_1}^4\,\bar{q_1}),\quad d=(1-\chi_1^4 \chi_2^4)(1-\bar{\chi_2}^4\bar{q_3})(1-\bar{\chi_1}^4\,\bar{q_2})
$$

Now the procedure is as follows. One treats the following two cases. Either all $a=b=c=d=0$ or
not all these coefficients vanish. 
 We will summarize our results here and give the details of the calculation in the appendix.

{\bf Classification Theorem.} The algebra $\BTheta$ is the {\em full}  matrix algebra {\em except} in the following cases in which it is
a proper subalgebra.

\begin{enumerate}
\item $q_1=q_2=q_3=1$ (the special bosonic cases)  and one of the following is true:

\begin{enumerate} 
\item All $\chi_i^2=1$ then $\BTheta$ is isomorphic to the commutative algebra in the case of no magnetic field above.

\item Two of the $\chi_i^4=-1$, the third one necessarily being equal to $1$.\end{enumerate}
\item If $q_i=-1$ (special fermionic cases)  and  $\chi_i^4=1$. This means that either 
\begin{enumerate}
\item all $\chi_i^2=-1$ or
\item  only one of the $\chi_i^2=-1$ the other two being $1$.
\end{enumerate}
\item $\bar q_1=q_2=q_3=\bar \chi^4_2$ and $\chi^2_1=1$ it follows that $\chi_2^4=\chi_3^4$. This is a one parameter family.
\item $q_1=q_2=q_3=\bar\chi_1^4$ and $\chi_2^2=1$ it follows that $ \chi_1^4=\bar \chi_3^4$. This is a one parameter family.
\item $q_1=q_2=\bar q_3=\bar \chi_1^4$ and $\chi_1^2=\bar\chi_2^2$. It follows that $\chi_3^4=1$. This is a one parameter family.

\end{enumerate}

 The subalgebra in the case (i)(b) is the most complicated. Notice that in this case $\Theta$ has integer entries and so  $\Ttheta\simeq \T^3=\T^3_{\Theta=0}$ is actually commutative, but
 $\BTheta$ is not. This can happen because we are looking at a sub-algebra of the non--commutative matrix algebra. 
  It is explicitly given as follows.  Consider $G_1=(1+U+V+W)(1+U^*+V^*+W^*)$ then the (2,2) entry of $\rho(G_1)$ will be of 
 the form $G_2=A-B+iC-iD$ where the $A,B,C,D$ and  polynomials in the $U,V,W,U^*,V^*,W^*$ of
 degree $0,1,2$ with positive integer coefficients. This is because the $\chi_i^2$ are $\pm i$ or $\pm 1$. Let $J$ be ideal of $\T^3$ 
 spanned by $G_1$ and $G_2$, let $J_{12}$ be the ideal spanned by $1+U^*+V^*+W^*$ and $J_{21}$ the ideal spanned by
 $1+U+V+W$. Then 
 \begin{equation}
 \label{special}
 \BTheta=\rho(\Ttheta)+\left(\begin{array}{cc}J&J_{12}\\ 
 J_{21}&J\end{array} \right)
\end{equation}
 
 The special fermionic case (ii) is related to Clifford algebras. Consider the quadratic form $Q$ on $\R^3$ with basis vectors $b_1,b_2,b_3$ given by $diag(\chi_1^2,\chi_2^2,\bar \chi_1^2\bar\chi_2^2)$. The condition $\chi_i^4=1$ translates to the fact that the entries are $\pm 1$.  Let $Cl={\it Cliff}(Q)\otimes {\mathbb C}$ be the complexified Clifford algebra of $Q$. In the fermionic case all the $q_i=-1$ so the generators of $\Ttheta$ anti--commute and there is a $C^*$ algebra map $\phi:\Ttheta\to Cl$ given by $\phi(U)=b_1,\phi(V)=b_2, \phi(W)=b_3$. Let ${\mathscr  J}:=ker(\phi)$ be the ideal defined by the kernel of $\phi$. 
Since the $\chi_i^2=\pm 1$ there is an involution $\hat{}:\Ttheta\to \Ttheta$ given by $\hat U=\chi_1^2U, \hat V=\chi_2^2 V$
and $\hat W=\bar \chi_1\bar \chi_2 W$. With these notations:
\begin{equation}
\label{cliffalg}
\BTheta=\{\left(\begin{array}{cc}a & b \\\hat b & \hat a\end{array}\right)+ J, \text{ with } a,b\in \Ttheta \text { and } J\in M_2({\mathscr J})\}.
\end{equation}

In the three families  the algebra $\BTheta$ is the $C^*$ algebra generated by
 $\Ttheta$ and two elements $A$ and $B$, which commute with each other and
$\Ttheta$, and satisfy equations $A^2=p$ and $B^2=q$ for fixed $p$ and $q$ in $\Ttheta$, i.e.\ there are adjoined square roots.
 For details on $p$ and $q$, see the Appendix.

\section{The Gyroid (G) case}
\begin{figure}
\begin{center}
\includegraphics[height=5cm]{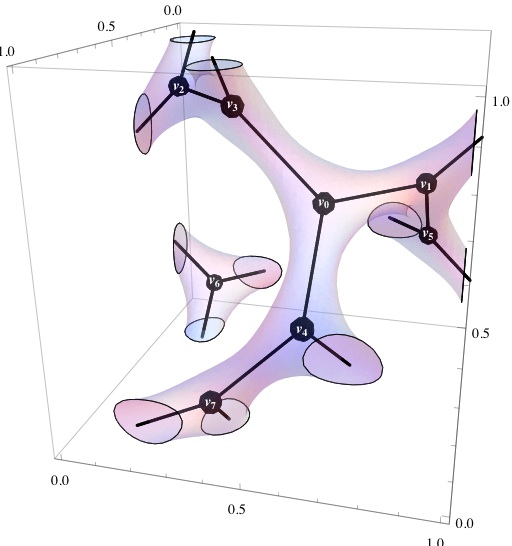}
\caption{One of the channels of the gyroid surface and its skeletal graph}
\label{Ggraph}
\end{center}
\end{figure}

We recall some of the setup from  \cite{KKWK}. The Gyroid and its graph are very complex and we will not give all the details here. 
One channel and the Gyroid graph  $\Gamma^+$  are shown in Figure \ref{Ggraph}. The symmetry group is $Ia\bar3d$. This means that the translation group is the bcc lattice. The graph $\bar \Gamma^+$ is the full square, see Figure \ref{graphfig}. 

We choose the generators of bcc to be the vectors  $g_1=\frac{1}{2}(1,-1,1),\; g_2=\frac{1}{2}(-1,1,1), \; g_3=\frac{1}{2}(1,1,-1)$.
These can be used these to fix the cocycle defining the interaction with the magnetic field:
$$ \theta_{12}=\frac{1}{2\pi}B\cdot (g_1\times g_2), \quad \theta_{13}=\frac{1}{2\pi}
 B\cdot (g_1\times g_3), \quad 
\theta_{23}=\frac{1}{2\pi}B\cdot (g_2\times g_3)
$$

The edge vectors of $\Gamma^+$ span the fcc group. Explicitly the edge vectors are

$e_1=\frac{1}{4} (-1,1,0),$
$e_2=\frac{1}{4} (0,-1,1),$
$e_3=\frac{1}{4} (1,0,-1),$
$e_4=\frac{1}{4} (1,1,0),$
$e_5=\frac{1}{4} (0,-1,-1),$
$e_6=\frac{1}{4} (-1,0,-1)$.

In the direct sum decomposition of $\H$ the Harper Hamiltonian reads

\begin{equation}
H_{\bgp}=\left(
\begin{array}{cccc}
0&U_1^*&U_2^*&U_3^*\\
U_1&0&U_6^*&U_5\\
U_2&U_6&0&U_4\\
U_3&U_5^*&U_4^*&0\\
\end{array}
\right)
\end{equation}

We choose the rooted spanning tree $\tau$ (root $A$, edges $e_1,e_2,e_3$)
Using this we obtain the following matrix Harper operator according to \cite{KKWK}

\begin{equation}
H_=\left(
\begin{array}{cccc}
0&1&1&1\\
1&0&U_1^*U_6^*U_2&U_1^*U_5U_3\\
1&U_2^*U_6U_1&0&U_2^*U_4U_3\\
1&U_3^*U_5^*U_1&U_3^*U_4^*U_2&0
\end{array}
\right)
=:\left(
\begin{array}{cccc}
0&1&1&1\\
1&0&A&B^*\\
1&A^*&0&C\\
1&B&C^*&0
\end{array}
\right)
\end{equation}

The operators $A,B,C$ again span a non--commutative three torus:  
\begin{equation}
AB=\alpha_1 BA, \quad AC=\bar\alpha_2CA, \quad BC=\alpha_3CB
\end{equation}
where now in terms of the $B$ field
$\alpha_1:=e^{2\pi i \theta_{12}},
\bar \alpha_2:= e^{2\pi i \theta_{13}}$,
$\alpha_3:=e^{2\pi i \theta_{23}}
$.

\subsection{The commutative case.} It is easy to check that generically the Hamiltonian has 4 distinct Eigenvalues. We can use
the character $\chi(A)=-1,\chi(b)=1,\chi(C)=-1$ for this. The corresponding Eigenvalues are $\pm \sqrt 5,\pm1$.
By the general theory we then know that the
commutative geometry if given by a generically unramified 4-fold cover of the three torus, see \cite{KKWK}.
The actual calculation of the branching behavior is more difficult. For this we have to analyze the characteristic 
polynomial of $H$ and thus we have to deal with a fourth order equation. Although it is in principle possible to solve the equation, this is
rather difficult and lengthy. 
We will treat this case in a subsequent paper \cite{KKWK3}. There we show that there are only 4 ramification points. 
This means that the locus is of real codimension 3 contrary to the D case where it was of codimension 2.
Furthermore the degenerations are 3 branches coming together at 2 points and 2 pairs of branches coming
together at the other two points.

\subsection{Non-commutative case}

To state the results of \cite{KKWK} we use
$$\phi_1=e^{\frac{\pi}{2} i \theta_{12}}, \quad \phi_2= e^{\frac{\pi}{2} i \theta_{31}},\quad
\phi_3= e^{\frac{\pi}{2} i \theta_{23}}, \quad \Phi=\phi_1\phi_2\phi_3
$$

{\bf Classification Theorem.}
\begin{enumerate}
\item If $\Phi\neq1$ or $\Phi=1$ and at least one $\alpha_i\neq 1$ and all $\phi_i$ are different then
$\BTheta=M_4(\T^3_{\Theta})$.

\item If $\phi_i=1$ for all $i$ then the algebra is the same as in the commutative case.

\item In all other cases $\B$ is non--commutative and $\BTheta\subsetneq M_4(\T^3_{\Theta})$. 
\end{enumerate}
Further information, which is too lengthy to reproduce here, about the case (iii) is in \cite{KKWK}.
We only wish to point out that the fermionic case $\alpha_i=-1$ is not a special case. 
Rather it is a mixed case in which two of the $\alpha_i=-1$ and one $\alpha_i=1$ which yields
a proper subalgebra involving 
a Clifford algebra.

\section{Conclusion}
We have treated all the triply periodic self-dual symmetric surface wire arrays ---given by the  P, D, G geometries---
using the methods developed in \cite{KKWK} to study their commutative and noncommutative geometry.
We gave  the commutative geometry as an explicit branched cover of the three torus
and classified all the noncommutative $C^*$ algebras that arise from turning on a constant magnetic field.

The G case was  considered before in \cite{KKWK}. As we discussed the P case can be reduced to information
contained in that paper as well.  Here we completely treated the D case
which has a much richer structure.  A new feature of the commutative case is that the branching
locus is  not of dimension zero, but rather of dimension one. A novel trait of the non--commutative case for the D surface
is the appearance of whole one--dimension families  where the algebra
drops to a proper subalgebra of the matrix algebra. 

An intriguing question is if these
two features are related.  Although the base space is $T^3$ in both cases, it parameterizes
completely different moduli. In the commutative case the parameters are the momenta, while
in the non--commutative case they are the parameters of the noncommutative torus which are
given by the magnetic field, which is completely absent in the commutative case. 
Thus there does not seem to be a direct relation, but  
one could expect such a relation on the grounds of a, yet to be determined, duality.
We leave this  for further investigation.

\section*{Acknowledgments}
RK thankfully acknowledges 
support from NSF DMS-0805881. BWK  thankfully acknowledges support from the  NSF under the grant PHY-0969689.

  Any opinions, findings and conclusions or 
recommendations expressed in this
 material are those of the authors and do not necessarily 
reflect the views of the National Science Foundation. 

\section*{Appendix}
In this appendix we give the details of the calculations for the D surface wire network.
As mentioned above the proof boils down to two major cases depending on wether the matrix
$X_3$ of equation (\ref{xmatrix}) is zero or not.

\subsection{The matrix $X_3 \neq 0$}
We also assume that all the $q_i\neq 1$. The case of all $q_i=1$ will be treated separately below.
The strategy is to reduce the matrix by conjugation so that only one term is non--zero.
After multiplication with the appropriate matrix one can obtain the matrix $E_{12}$ and hence the whole
matrix algebra.

The subcases one treats are (I) $a\neq 0$ and (II) $a=0$. In case (I), one can successively kill all the entries except for
the one proportional to $1$. Explicitly,
after performing the three operations $X_4= \bar{q_1}X_3 -\chi_1^4 \rho(U_{f_2}) X_3 \rho(U_{f_2}^*)$, then $X_5= q_1X_4 -\chi_2^4 \rho(U_{f_3}) X_4\rho(U_{f_3}^*)$, and finally $X_6= \bar{q_2}X_5 -\bar{\chi_1}^4 \bar{\chi_2}^4 \rho(U_{f_4}) X_5 \rho(U_{f_4}^*)$, we obtain $X_6=a''E_{12}$ which has only one possibly non--zero entry,
$$
a^{\prime\prime}=(\bar{q_2}-1)(q_1-1)(\bar{q_1}-1)(1-\chi_1^4 \chi_2^4)(1-\bar{\chi_2}^4)(1-\bar{\chi_1}^4) 
$$
Hence $X_6$ can be brought to $E_{12}$ by dividing by $a^{\prime\prime}$, provided it is non--zero. Since we assume not all $q_i=1$ and $a\neq0$, the remaining cases 
are when one or both $q_1=1,q_2=1$ but not all three $q_i=1$. These can be handled similarly and all lead to the full 
matrix algebra.

The case (II) splits as several subcases corresponding to the factors of $a$: (A) $\chi_1^4=1$, (B) $\chi_2^4=1$ and (C) $\chi_1^4\chi_2^4=1$.  All these cases are similar, we show how to treat (A). In this case, we already know that $b=0$ and
if we further assume that $d=0$ it follows $c=0$ and we are in the case $X_3=0$. So, we assume $d\neq0$. If $c=0$ there is only one term and we are done.  If $c\neq0$ then we can conjugate with $\rho(V)$ and kill the $V^*$ term leaving only the $W^*$ term
and we are done.

\subsection{The matrix $X_3=0$.}
This is more tedious. The cases we get from assuming that all the coefficients are zero are:
(A) $\chi_1^4=\chi_2^4=1$ which implies $q_1=\bar q_2=q_3$. 
(B) $\chi_1^4=1$ and 
(1) $q_3=\bar\chi_2^4$ which implies $\bar q_1= q_2=q_3$, $\chi_2^4=\chi_3^4$ or 
(2) $q_2=1$ which implies $q_1=q_2=1$, $\bar \chi_2^4=\chi_3^4$.
(C) $\chi_2^4=1$ and 
(1) $q_2=\bar \chi_1^4$ which implies $q_1=q_2=q_3$, $\bar\chi_1^4=\chi_3^4$ or 
(2) $q_3=1$ which implies $q_1=q_3=1$, $\chi_1^4=\chi_3^4$. 
And finally (D) $\chi_1^4=\bar \chi_2^4=1$ and 
(1) $q_1=\chi_2^4$ which implies  $q_1=q_2=\bar q_3$, $\chi_3^4=1$ or 
(2) $q_2=1$ which implies $q_2=q_3=1$, $q_1=\chi_3^4$. 

Again all $q_i=1$ will be treated separately.

In case (A), either $q_3\neq \bar q_3$ and we can proceed as usual and obtain the full matrix algebra. Or $q_3=\bar q_3$, and then either all $q_i=1$ or  all $q_i=-1$. In the latter case we will show that
$\BTheta$ is indeed the algebra given  by (\ref{cliffalg}). For the time being denote that algebra by $\B'$.
It is easy to check that $\B'$ is a subalgebra. It is also proper, since it does not surject onto the image of $\phi$, for
instance $E_{12}$ is not in the image. Since 
$$H=\left(\begin{array}{cc}0 & 1+\hat U +\hat V+\hat W \\1+U+V+W & 0\end{array}\right)+\left(\begin{array}{cc}0 & U^*-\hat U +V^*-\hat V+W^*-\hat W \\0 & 0\end{array}\right)$$
we see that $H\in \B'$, likewise
one checks that $\rho(\Ttheta)\subset \B'$ and hence $\BTheta\subset \B'$. To get the other inclusion,
one proceeds in the usual fashion to obtain the matrices 
$$
I=\left(\begin{array}{cc}0 & 1\\1 & 0\end{array}\right),
\left(\begin{array}{cc}0 & U^*\\U&0\end{array}\right),
\left(\begin{array}{cc}0 & V^* \\V & 0\end{array}\right),
\left(\begin{array}{cc}0 & W^* \\W & 0\end{array}\right)
$$
By multiplying $I$ with elements of $\rho(\Ttheta)$ and subtracting we get the matrices $U^*-\hat U E_{12}$,$U^*-\hat U E_{12}$,$V^*-\hat V E_{12}$ and $W^*-\hat W E_{12}$ which generate $\mathscr J$. Thus $\B'\subset \BTheta$.

In case (B) (1) with the assumption $q_i\neq 1$, we can either have $\chi_1^2\neq1$ in which case the usual procedure
produces the full matrix algebra or $\chi^2_1=1$ in which case we obtain the matrices 
$
A=\left(\begin{array}{cc}0 & U^*\\1&0\end{array}\right),
C=\left(\begin{array}{cc}0 & W^* \\V & 0\end{array}\right)
$
and their adjoints.   Set 
$B=C\rho(V^*)=\left(\begin{array}{cc}0 & \bar\chi_2^2W^*V^* \\1 & 0\end{array}\right)$. Then both $A$ and $B$ commute
with $\rho(\Ttheta)$ and with each other. Now $A^2=\rho(U)$ and $B^2=\chi_2^2\rho(W^*V^*)$. Since $H=A+A^*+C+C^*$
we see that $H$ is in the $C^*$ sub--algebra spanned by $\rho(\Ttheta)$, $A$ and $B$ with the given relations.
 To show that this is not the full matrix algebra, we can use the mapping of $\phi:\Ttheta\to \T^2_{\frac{1}{2}}$ given
by $\phi(U)=S, \phi(V)=T, \phi(W)=S^*T$ where $S,T$ are the generators of $\T_{\frac{1}{2}}^2$, which satisfy $ST=-TS$.
We see that $ker(\phi)$ is the two sided $C^*$ ideal generated by $V^*W-U$. The map $\phi$ induced a 
map $\hat\phi:M_2(\Ttheta)\to M_2(\T^2_{\frac{1}{2}})$. Since the image of $A$ is the image of  $\bar\chi_2^2 B$,
we see that the image of $\B$ is generated by $\hat\phi\rho(\Ttheta)$ and $\hat\phi(A)$, which does not contain $E_{12}$.
Hence $\hat\phi|_{\BTheta}$ is not surjective and $\BTheta$ is not the full algebra.  From this it is also easy to see that in $M_2(\Ttheta)$, $A$ and $B$ satisfy no other relations modulo $\rho(\Ttheta)$. This is the family (iii).

The case (B)(2) yields the full algebra unless $q_3\neq 1$ and hence all $q_i=1$.

The case (C) is completely analogous upon switching $U$ and $V$. (C)(1) yields the family $(iv)$.

In the case (D) $W$ plays the special role, which $U$ played in (B)(1) and hence the condition is that $\chi_1^2=\chi_2^2=1$. This yields the case of  the family (v).

\end{document}